\documentclass[12pt,a4paper]{article}
\usepackage{xcolor}
\usepackage{ifthen} % for conditional statements
\usepackage{booktabs}

\newboolean{pdflatex}
\setboolean{pdflatex}{true} % False for eps figures 

\newboolean{articletitles}
\setboolean{articletitles}{true} % False removes titles in references

\newboolean{uprightparticles}
\setboolean{uprightparticles}{false} %True for upright particle symbols

\newboolean{inbibliography}
\setboolean{inbibliography}{false} %True once you enter the bibliography

\usepackage[top=1in, bottom=1.25in, left=1in, right=1in]{geometry}

\usepackage{microtype}
\usepackage{lineno}  % for line numbering during review
\usepackage{xspace} % To avoid problems with missing or double spaces after
\usepackage{caption} %these three command get the figure and table captions automatically small

\usepackage{graphicx}  % to include figures (can also use other packages)
\usepackage{color}
\usepackage{colortbl}
\graphicspath{{./figs/}} % Make Latex search fig subdir for figures

\usepackage{amsmath} % Adds a large collection of math symbols
\usepackage{amssymb}
\usepackage{amsfonts}
\usepackage{upgreek} % Adds in support for greek letters in roman typeset

\newcommand*\patchAmsMathEnvironmentForLineno[1]{%
\expandafter\let\csname old#1\expandafter\endcsname\csname #1\endcsname
\expandafter\let\csname oldend#1\expandafter\endcsname\csname
end#1\endcsname
 \renewenvironment{#1}%
   {\linenomath\csname old#1\endcsname}%
   {\csname oldend#1\endcsname\endlinenomath}%
}
\newcommand*\patchBothAmsMathEnvironmentsForLineno[1]{%
  \patchAmsMathEnvironmentForLineno{#1}%
  \patchAmsMathEnvironmentForLineno{#1*}%
}
\AtBeginDocument{%
\patchBothAmsMathEnvironmentsForLineno{equation}%
\patchBothAmsMathEnvironmentsForLineno{align}%
\patchBothAmsMathEnvironmentsForLineno{flalign}%
\patchBothAmsMathEnvironmentsForLineno{alignat}%
\patchBothAmsMathEnvironmentsForLineno{gather}%
\patchBothAmsMathEnvironmentsForLineno{multline}%
\patchBothAmsMathEnvironmentsForLineno{eqnarray}%
}

\usepackage{hyperref}    % Hyperlinks in references
\usepackage[all]{hypcap} % Internal hyperlinks to floats.

\usepackage{xspace} 
\usepackage{upgreek}

\def\lhcb {\mbox{LHCb}\xspace}

\def\MagUp {\mbox{\em Mag\kern -0.05em Up}\xspace}

\ifthenelse{\boolean{uprightparticles}}%
{

 \def\Ppi         {\ensuremath{\uppi}\xspace}

 \def\PDelta      {\ensuremath{\Delta}\xspace}                 
 \def\PXi      {\ensuremath{\Xi}\xspace}                 
 \def\PLambda      {\ensuremath{\Lambda}\xspace}                 
 \def\PSigma      {\ensuremath{\Sigma}\xspace}                 
 \def\POmega      {\ensuremath{\Omega}\xspace}                 
 \def\PUpsilon      {\ensuremath{\Upsilon}\xspace}                 
 
 \def\PB      {\ensuremath{\mathrm{B}}\xspace}                 
                  
 \def\PD      {\ensuremath{\mathrm{D}}\xspace}

 \def\PK      {\ensuremath{\mathrm{K}}\xspace}

 \def\Pb      {\ensuremath{\mathrm{b}}\xspace}                 
 \def\Pc      {\ensuremath{\mathrm{c}}\xspace}

 \def\Pi      {\ensuremath{\mathrm{i}}\xspace}

 \def\Ps      {\ensuremath{\mathrm{s}}\xspace}

}
{

 \def\Ppi         {\ensuremath{\pi}\xspace}

 \mathchardef\PDelta="7101
 \mathchardef\PXi="7104
 \mathchardef\PLambda="7103
 \mathchardef\PSigma="7106
 \mathchardef\POmega="710A
 \mathchardef\PUpsilon="7107
                  
 \def\PB      {\ensuremath{B}\xspace}                 
                  
 \def\PD      {\ensuremath{D}\xspace}

 \def\PK      {\ensuremath{K}\xspace}

 \def\Pb      {\ensuremath{b}\xspace}                 
 \def\Pc      {\ensuremath{c}\xspace}

 \def\Pi      {\ensuremath{i}\xspace}

 \def\Ps      {\ensuremath{s}\xspace}

}

\makeatletter
\ifcase \@ptsize \relax% 10pt
  \newcommand{\miniscule}{\@setfontsize\miniscule{4}{5}}% \tiny: 5/6
\or% 11pt
  \newcommand{\miniscule}{\@setfontsize\miniscule{5}{6}}% \tiny: 6/7
\or% 12pt
  \newcommand{\miniscule}{\@setfontsize\miniscule{5}{6}}% \tiny: 6/7
\fi
\makeatother

\DeclareRobustCommand{\optbar}[1]{\shortstack{{\miniscule (\rule[.5ex]{1.25em}{.18mm})}
  \\ [-.7ex] $#1$}}

\def\squark    {{\ensuremath{\Ps}}\xspace}

\def\cquark    {{\ensuremath{\Pc}}\xspace}

\def\bquark    {{\ensuremath{\Pb}}\xspace}

\def\pion   {{\ensuremath{\Ppi}}\xspace}

\def\pim    {{\ensuremath{\pion^-}}\xspace}

\def\kaon    {{\ensuremath{\PK}}\xspace}
  \def\Kbar    {{\kern 0.2em\overline{\kern -0.2em \PK}{}}\xspace}

\def\KorKbar    {\kern 0.18em\optbar{\kern -0.18em K}{}\xspace}
\def\Kz      {{\ensuremath{\kaon^0}}\xspace}

  \def\Dbar    {{\kern 0.2em\overline{\kern -0.2em \PD}{}}\xspace}
\def\D       {{\ensuremath{\PD}}\xspace}

\def\DorDbar    {\kern 0.18em\optbar{\kern -0.18em D}{}\xspace}
\def\Dz      {{\ensuremath{\D^0}}\xspace}

\def\Dm      {{\ensuremath{\D^-}}\xspace}

\def\Dstarm  {{\ensuremath{\D^{*-}}}\xspace}

\def\Dsp     {{\ensuremath{\D^+_\squark}}\xspace}
\def\Dsm     {{\ensuremath{\D^-_\squark}}\xspace}

\def\Dssm    {{\ensuremath{\D^{*-}_\squark}}\xspace}

\def\B       {{\ensuremath{\PB}}\xspace}
\def\Bbar    {{\ensuremath{\kern 0.18em\overline{\kern -0.18em \PB}{}}}\xspace}

\def\BorBbar    {\kern 0.18em\optbar{\kern -0.18em B}{}\xspace}
\def\Bz      {{\ensuremath{\B^0}}\xspace}

\def\Bu      {{\ensuremath{\B^+}}\xspace}

\def\Bp      {{\ensuremath{\Bu}}\xspace}

\def\Bs      {{\ensuremath{\B^0_\squark}}\xspace}
\def\Bsb     {{\ensuremath{\Bbar{}^0_\squark}}\xspace}

\def\Bc      {{\ensuremath{\B_\cquark^+}}\xspace}

  \def\Y#1S{\ensuremath{\PUpsilon{(#1S)}}\xspace}% no space before {...}!

\def\Lz          {{\ensuremath{\PLambda}}\xspace}
\def\Lbar        {{\ensuremath{\kern 0.1em\overline{\kern -0.1em\PLambda}}}\xspace}
\def\LorLbar    {\kern 0.18em\optbar{\kern -0.18em \PLambda}{}\xspace}

\def\Lb      {{\ensuremath{\Lz^0_\bquark}}\xspace}

\def\Lc      {{\ensuremath{\Lz^+_\cquark}}\xspace}

\def\to                 {\ensuremath{\rightarrow}\xspace}

\def\AT#1     {\ensuremath{A_{\mathrm{T}}^{#1}}\xspace}           % 2

\def\C#1      {\ensuremath{\mathcal{C}_{#1}}\xspace}                       % 9
\def\Cp#1     {\ensuremath{\mathcal{C}_{#1}^{'}}\xspace}                    % 7
\def\Ceff#1   {\ensuremath{\mathcal{C}_{#1}^{\mathrm{(eff)}}}\xspace}        % 9  
\def\Cpeff#1  {\ensuremath{\mathcal{C}_{#1}^{'\mathrm{(eff)}}}\xspace}       % 7
\def\Ope#1    {\ensuremath{\mathcal{O}_{#1}}\xspace}                       % 2
\def\Opep#1   {\ensuremath{\mathcal{O}_{#1}^{'}}\xspace}                    % 7

\newcommand{\tev}{\ifthenelse{\boolean{inbibliography}}{\ensuremath{~T\kern -0.05em eV}}{\ensuremath{\mathrm{\,Te\kern -0.1em V}}}\xspace}
\newcommand{\gev}{\ensuremath{\mathrm{\,Ge\kern -0.1em V}}\xspace}
\newcommand{\mev}{\ensuremath{\mathrm{\,Me\kern -0.1em V}}\xspace}
\newcommand{\kev}{\ensuremath{\mathrm{\,ke\kern -0.1em V}}\xspace}
\newcommand{\ev}{\ensuremath{\mathrm{\,e\kern -0.1em V}}\xspace}
\newcommand{\gevc}{\ensuremath{{\mathrm{\,Ge\kern -0.1em V\!/}c}}\xspace}
\newcommand{\mevc}{\ensuremath{{\mathrm{\,Me\kern -0.1em V\!/}c}}\xspace}
\newcommand{\gevcc}{\ensuremath{{\mathrm{\,Ge\kern -0.1em V\!/}c^2}}\xspace}
\newcommand{\gevgevcccc}{\ensuremath{{\mathrm{\,Ge\kern -0.1em V^2\!/}c^4}}\xspace}
\newcommand{\mevcc}{\ensuremath{{\mathrm{\,Me\kern -0.1em V\!/}c^2}}\xspace}

\def\invfb   {\ensuremath{\mbox{\,fb}^{-1}}\xspace}

\def\invps{\ensuremath{{\mathrm{ \,ps^{-1}}}}\xspace}

\newcommand{\stat}{\ensuremath{\mathrm{\,(stat)}}\xspace}
\newcommand{\syst}{\ensuremath{\mathrm{\,(syst)}}\xspace}

\def\gsim{{~\raise.15em\hbox{$>$}\kern-.85em
          \lower.35em\hbox{$\sim$}~}\xspace}
\def\lsim{{~\raise.15em\hbox{$<$}\kern-.85em
          \lower.35em\hbox{$\sim$}~}\xspace}

\def\pt         {\mbox{$p_{\mathrm{ T}}$}\xspace}

\def\tell1  {TELL1\xspace}
\def\ukl1   {UKL1\xspace}

\usepackage{cite} % Allows for ranges in citations
\usepackage{mciteplus}

\usepackage{longtable} % only for template; not usually to be used in PAPERs

\usepackage{overpic}
\usepackage{etoolbox}

\def\DSSm    {{\ensuremath{\D^{\scalebox{0.4}{(}*\scalebox{0.4}{)}-}_{\squark}}}\xspace}
\def\DSm    {{\ensuremath{\D^{\scalebox{0.4}{(}*\scalebox{0.4}{)}-}}}\xspace}
\def\DSSM    {{\ensuremath{\D^{\scalebox{0.4}{(}*\scalebox{0.4}{)}-}_{(\squark)}}}\xspace}
\def\DSM    {{\ensuremath{\D^{\scalebox{0.4}{}\scalebox{0.4}{}-}_{(\squark)}}}\xspace}
\def\DSP    {{\ensuremath{\D^{\scalebox{0.4}{}\scalebox{0.4}{}+}_{(\squark)}}}\xspace}

\robustify{\DSm}
\robustify{\DSSm}
\robustify{\DSSM}
\robustify{\DSM}
\robustify{\DSP}

\newcommand{\taufs}{\ensuremath{\tau^{\rm fs}_{\Bs}}\xspace}

\newcommand{\dB}{\ensuremath{\rm -0.0115} }
\newcommand{\eStatdB}{\ensuremath{0.0053}}

\newcommand{\eSystdB}{\ensuremath{0.0041}}

\newcommand{\tB}{\ensuremath{\rm 1.547} }
\newcommand{\eStattB}{\ensuremath{0.013} } % 0.0116
\newcommand{\eSysttB}{\ensuremath{0.010} } % 0.00947
\newcommand{\eReftB}{\ensuremath{0.004} }
\newcommand{\refB}{\ensuremath{\,(\tau_{\it B})}} 

\newcommand{\dD}{\ensuremath{1.0131} }
\newcommand{\eStatdD}{\ensuremath{0.0117} }

\newcommand{\eSystdD}{\ensuremath{0.0065} }
 
\newcommand{\tD}{\ensuremath{0.5064} } 
\newcommand{\eStattD}{\ensuremath{0.0030} }

\newcommand{\eSysttD}{\ensuremath{0.0017} }
\newcommand{\eReftD}{\ensuremath{0.0017} }
\newcommand{\refD}{\ensuremath{\,(\tau_{\it D})}} 

\begin{document}

%%%%%%%%%%%%%%%%%%%%%%%%%
%%%%% Title     %%%%%%%%%
%%%%%%%%%%%%%%%%%%%%%%%%%
\renewcommand{\thefootnote}{\fnsymbol{footnote}}
\setcounter{footnote}{1}

% %%%%%%% CHOOSE TITLE PAGE--------
%\onecolumn
\begin{titlepage}
\pagenumbering{roman}

% Header ---------------------------------------------------
\vspace*{-1.5cm}
\centerline{\large EUROPEAN ORGANIZATION FOR NUCLEAR RESEARCH (CERN)}
\vspace*{1.5cm}
\noindent
\begin{tabular*}{\linewidth}{lc@{\extracolsep{\fill}}r@{\extracolsep{0pt}}}
%\ifthenelse{\boolean{pdflatex}}% Logo format choice
%{\vspace*{-2.7cm}\mbox{\!\!\!\includegraphics[width=.14\textwidth]{lhcb-logo.pdf}} & &}%
%{\vspace*{-1.2cm}\mbox{\!\!\!\includegraphics[width=.12\textwidth]{lhcb-logo.eps}} & &}%
\\
 & & CERN-EP-2017-070 \\  % ID 
 & & LHCb-PAPER-2017-004 \\  % ID 
 & &  September, 8, 2017 \\%\today \\ % Date - Can also hardwire e.g.: 23 March 2010
 & & \\
% not in paper \hline
\end{tabular*}

%\vspace*{4.0cm}
\vspace*{2.0cm}

% Title --------------------------------------------------
{\normalfont\bfseries\boldmath\huge
\begin{center}
 Measurement of \Bs and \Dsm meson lifetimes
\end{center}
}

\vspace*{1.0cm}

% Authors -------------------------------------------------
\begin{center}
%In the footnote, replace 'paper' by 'Letter' in case of submission to PRL or PLB 
The LHCb collaboration\footnote{Authors are listed at the end of this paper.}
\end{center}

% Abstract -----------------------------------------------
\begin{abstract}
\noindent
We report on a measurement of the flavor-specific \Bs lifetime and of the \Dsm lifetime using proton-proton collisions at center-of-mass energies of 7 and 8 TeV, collected by the LHCb experiment and corresponding to 3.0 fb$^{-1}$ of integrated luminosity. Approximately 407\,000  $\Bs \to \DSSm \mu^+\nu_\mu $ decays are partially reconstructed in the  $K^+K^-\pi^-\mu^+$ final state. The \Bs and \Dsm natural widths are determined using, as a reference,  kinematically similar $\Bz \to \DSm \mu^+\nu_\mu$ decays reconstructed in the same final state. The resulting differences between widths of \Bs and \Bz mesons and of \Dsm and \Dm mesons are $\Delta_\Gamma(B) = \dB\pm \eStatdB\stat \pm \eSystdB\syst$\invps and $\Delta_\Gamma(D) = \dD\pm \eStatdD\stat \pm \eSystdD\syst$\invps, respectively.  Combined with the known \Bz and \Dm lifetimes, these yield the flavor-specific \Bs lifetime, $\taufs =  \tB \pm \eStattB\stat \pm \eSysttB\syst\pm\eReftB \refB$\,ps  and the \Dsm lifetime, $\tau_{\Dsm} = \tD \pm \eStattD\stat \pm \eSysttD\syst\pm \eReftD \refD$\,ps. The last uncertainties originate from the limited knowledge of the \Bz and \Dm lifetimes. The results improve upon current determinations.
\end{abstract}

\vspace*{1.0cm}

\begin{center}
  Published in Phys.~Rev.~Lett. 119, 101801 (2017).
\end{center}

\vspace{\fill}

{\footnotesize 
\centerline{\copyright~CERN on behalf of the \lhcb collaboration, licence \href{http://creativecommons.org/licenses/by/4.0/}{CC-BY-4.0}.}}
\vspace*{2mm}

\end{titlepage}

%%%%%%%%%%%%%%%%%%%%%%%%%%%%%%%%
%%%%%  EOD OF TITLE PAGE  %%%%%%
%%%%%%%%%%%%%%%%%%%%%%%%%%%%%%%%

%  empty page follows the title page ----
\newpage
\setcounter{page}{2}
\mbox{~}
%\newpage
%
%% Author List ----------------------------
%%  You need to get a new author list!
%\input{LHCb_authorlist.tex}
%
%The author list for journal publications is provided by the Membership Committee shortly after 'approval to go to paper' has been given.
%%It will be made available on the page
%%\verb!http://www.physik.uzh.ch/~strauman/forMemCo/LHCb-PAPER-XXXX-XXX/! .
%It will be sent to you by email shortly after a paper number has beens assigned.
%The author list should be included already at first circulation, 
%to allow new members of the collaboration to verify whether they have been included correctly.
%Occasionally a misspelled name is corrected or associated institutions become full members.
%In that case, a new author list will be sent to you.
%In case line numbering doesn't work well after including the authorlist, try moving the \verb!\bigskip! after the last author to a separate line.
%
%
%The authorship for Conference Reports should be ``The LHCb
%  collaboration'', with a footnote giving the name(s) of the contact
%  author(s), but without the full list of collaboration names.

\cleardoublepage

%\twocolumn
% %%%%%%%%%%%%% ---------

\renewcommand{\thefootnote}{\arabic{footnote}}
\setcounter{footnote}{0}

%%%%%%%%%%%%%%%%%%%%%%%%%
%%%%% Main text %%%%%%%%%
%%%%%%%%%%%%%%%%%%%%%%%%%

\pagestyle{plain} % restore page numbers for the main text
\setcounter{page}{1}
\pagenumbering{arabic}

%% Uncomment during review phase. 
%% Comment before a final submission.
%\linenumbers

\noindent
Comparisons of precise measurements and predictions associated with quark-flavor dynamics probe the existence of unknown particles at energies much higher than those directly accessible at particle colliders. The precision of the predictions is often limited by the strong-interaction theory at low energies, where calculations are intractable. Predictive power is recovered by resorting to effective models such as heavy-quark expansion~\cite{Lenz:2014jha}
%~\cite{Shifman:1986sm, Shifman:1987rj, Eichten:1989zv, Guberina:1979fe, Guberina:1979re, Isgur:1989vq, Isgur:1989ed, Grinstein:1990mj, Falk:1990yz, Georgi:1990um}, 
which rely on an expansion of the quantum chromodynamics corrections in powers of $1/m$, where $m$ is the mass of the heavy quark in a bound system of a heavy quark and a light quark. These predictions are validated and refined using lifetime measurements of heavy hadrons. Hence, improved lifetime measurements ultimately enhance the reach in searches for non-standard-model physics. Currently, more precise measurements are particularly important as predictions of the lifetime ratio between \Bs and \Bz mesons show a 2.5 standard-deviation discrepancy from measurements. \par Measurements of the ``flavor-specific" \Bs meson lifetime, $\taufs$, have additional relevance. This empirical quantity is a function of the natural widths of the two mass eigenstates resulting from \Bs--\Bsb oscillations, and therefore allows an indirect determination of the width difference that can be compared with direct determinations in tests for non-standard-model physics~\cite{HFAG}. The lifetime $\taufs$ is measured with a single-exponential fit to the distribution of decay time in final states to which only one of \Bs and \Bsb mesons can decay~\cite{Hartkorn:1999ga}. The current best determination, $\taufs = 1.535 \pm 0.015 ({\rm stat}) \pm 0.014({\rm syst}) $\,ps~\cite{LHCb-PAPER-2014-037}, obtained by the LHCb collaboration using hadronic $\Bs \to \Dsm \pi^+$ decays, has similar statistical and systematic uncertainties.  Semileptonic \Bs decays, owing to larger signal yields than in hadronic decays, offer richer potential for precise $\taufs$ measurements. However, neutrinos and low-momentum neutral final-state particles prevent the full reconstruction of such decays. This introduces systematic limitations associated with poor knowledge of backgrounds and difficulties in obtaining the decay time from the observed decay-length distribution. Indeed, the result $\taufs = 1.479 \pm 0.010\stat  \pm 0.021\syst $\,ps~\cite{Abazov:2014rua}, based on a $\Bs \to \DSSm \mu^+\nu_\mu X$ sample from the D0 collaboration,  is limited by the systematic uncertainty. Throughout this Letter, the symbol $X$ identifies any decay product, other than neutrinos, not included in the candidate reconstruction, and the inclusion of charge-conjugate processes is implied.

In this Letter, we use a novel approach that suppresses the above limitations and achieves a precise measurement of the flavor-specific \Bs meson lifetime. The lifetime is determined from the variation in the \Bs signal yield as a function of decay time, relative to that of \Bz decays that are reconstructed in the same final state and whose lifetime is precisely known. The use of kinematically similar \Bz decays as a reference allows the reduction of the uncertainties from partial reconstruction and lifetime-biasing selection criteria. The analysis also yields a significantly improved determination of the \Dsm lifetime over the current best result, $\tau_{\Dsm}= 0.5074 \pm 0.0055\stat \pm 0.0051\syst$ ps, reported by the FOCUS collaboration~\cite{Link:2005ew}. 

We analyze proton-proton collisions at center-of-mass energies of 7 and 8 TeV collected by the LHCb experiment in 2011 and 2012 and corresponding to an integrated luminosity of 3.0 fb$^{-1}$. We use a sample of approximately 407\,000 $\Bs \to \Dssm \mu^+\nu_\mu$ and $\Bs \to \Dsm \mu^+\nu_\mu$ ``signal" decays, and a sample of approximately 108\,000 $\Bz \to \Dstarm \mu^+\nu_\mu$ and $\Bz \to \Dm \mu^+\nu_\mu$ ``reference" decays.  The $D$ candidates are reconstructed as combinations of $K^+$, $K^-$, and $\pi^-$ candidates originating from a common vertex, displaced from any proton-proton interaction vertex. The $B^0_{(s)}$ candidates, $K^+ K^- \pi^- \mu^+$,  are formed by $D$ candidates associated with muon candidates originating from another common displaced vertex.  We collectively refer to the signal and reference decays as $\Bs \to [K^+K^-\pi^-]_\DSSm \mu^+\nu_\mu$ and $\Bz \to [K^+K^-\pi^-]_\DSm \mu^+\nu_\mu$, respectively. A fit to the ratio of event yields between the signal and reference  decays as a function of $B^0_{(\squark)}$ decay time, $t$,  determines $\Delta_\Gamma(B) \equiv 1/\taufs - \Gamma_d$, where $\Gamma_d$ is the known natural width of the \Bz meson. A similar fit performed as a function of the $D^-_{(s)}$ decay time determines the decay-width difference between \Dsm and \Dm mesons, $\Delta_\Gamma(D)$. Event yields are determined by fitting the ``corrected-mass" distribution of the candidates, $m_{\rm corr} = p_{\perp, D\mu} + \sqrt{m^2_{D\mu}+p^2_{\perp, D\mu}}$~\cite{Kodama:1991ij}. This is determined from the invariant mass of the $D_{(s)}^-\mu^+$ pair, $m_{D\mu}$, and the component of its momentum perpendicular to the $B^0_{(s)}$ flight direction, $p_{\perp, D\mu}$, to compensate for the average momentum of unreconstructed decay products. The flight direction is the line connecting the $B^0_{(s)}$ production and decay vertices; the decay time $t = m_B L k/ p_{D\mu}$ uses the known $B^0_{(s)}$ mass, $m_B$~\cite{PDG2016}, the measured $B^0_{(s)}$ decay length, $L$,  and the momentum of the $D^-_{(s)}\mu^+$ pair, $p_{D\mu}$. The scale factor $k$ corrects $p_{D\mu}$ for the average momentum fraction carried by decay products excluded from the reconstruction~\cite{Abulencia:2006ze, Leonardo:2006fq}. The effects of decay-time acceptances and resolutions, determined from simulation, are included.

The LHCb detector is a single-arm forward spectrometer equipped with precise charged-particle vertexing and tracking detectors, hadron-identification detectors, calorimeters, and muon detectors, optimized for the study of bottom- and  charm-hadron decays~\cite{Alves:2008zz,Aaij:2014jba}. Simulation~\cite{LHCb-PROC-2011-006, LHCb-PROC-2010-056} is used to identify all relevant sources of bottom-hadron decays, model the mass distributions, and correct for the effects of incomplete kinematic reconstructions, relative decay-time acceptances, and decay-time resolutions. The unknown details of the \Bs decay dynamics are modeled in the simulation through empirical form-factor parameters~\cite{Caprini:1997mu}, assuming values inspired by the known $B^0$ form factors~\cite{HFAG}. We assess the impact of these assumptions on the systematic uncertainties. 

The online selection requires a muon candidate, with transverse momentum exceeding 1.5--1.8\gevc, associated with one, two, or three charged particles, all with origins displaced from the proton-proton interaction points~\cite{Aaij:2012me}. In the offline reconstruction, the muon is combined with charged particles consistent with the topology and kinematics of signal $\Bs \to [K^+K^-\pi^-]_\DSSm \mu^+\nu_\mu$ and reference  $\Bz \to [K^+K^-\pi^-]_\DSm \mu^+\nu_\mu$ decays.  The range of $K^+K^-\pi^-$ mass is restricted around the known values of the \DSM meson masses such that cross-contamination between signal and reference samples is smaller than 0.1\%, as estimated from simulation.  We also reconstruct ``same-sign" $K^+K^-\pi^-\mu^-$ candidates, formed by charm and muon candidates with same-sign charge, to model combinatorial background from accidental $D^{-}_{(s)} \mu^+$ associations. The event selection is optimized toward suppressing the background under the charm signals and making same-sign candidates a reliable model for the combinatorial background: track- and vertex-quality, vertex-displacement, transverse-momentum, and particle-identification criteria are chosen to minimize shape and yield differences between same-sign and signal candidates in the $m_{D\mu} > 5.5\gevcc$ region, where genuine bottom-hadron decays are kinematically excluded and combinatorial background dominates.  Mass vetoes suppress background from misreconstructed decays such as $\Bs \to \psi^{(')}(\to \mu^+\mu^-)\phi (\to K^+K^-)$ decays where a muon is misidentified as a pion, $\Lb \to \Lc (\to pK^-\pi^+) \mu^- \bar{\nu}_\mu X$ decays where the proton is misidentified as a kaon or a pion, and $B^0_{(s)} \to D^-_{(s)}\pi^+$ decays where the pion is misidentified as a muon. Significant contributions arise from decays of a bottom hadron into pairs of charm hadrons, one peaking at the $D^-_{(s)}$ mass and the other decaying semileptonically, or into single charm hadrons and other particles.  Such decays include $\B^0_{(\squark)} \to 
\D^{\scalebox{0.4}{(}*\scalebox{0.4}{)}-}_{(\squark)}\DSP$,  $\Bp \to \Dbar{}^{\scalebox{0.4}{(}*\scalebox{0.4}{)}0} D^{\scalebox{0.4}{(}*\scalebox{0.4}{)}+}$, $\Bp \to \Dm \mu^+ \nu_\mu X$, $\Bp \to \DSSm K^+ \mu^+ \nu_\mu X$, $\Bz \to \DSSm \Kz \mu^+ \nu_\mu X$, $\Bs \to \Dz\Dsm K^+$, $\Bs \to \Dm \Dsp \Kz$, $\Lb \to \Lc \DSSm X$, and $\Lb \to \Dsp \Lz \mu^- \bar{\nu}_\mu X$ decays. We suppress these backgrounds with a threshold, linearly dependent on $m_{\rm corr}$, applied to the $\D^-_{(s)}$ momentum component perpendicular to the $\B^0_{(\squark)}$ flight direction. Finally, a $t>0.1$\,ps requirement on the $\D^-_{(\squark)}$ proper decay time renders the signal- and reference-decay acceptances as functions of decay time more similar, with little signal loss.

\begin{figure}
\centering
\begin{overpic}[width=0.48\textwidth]{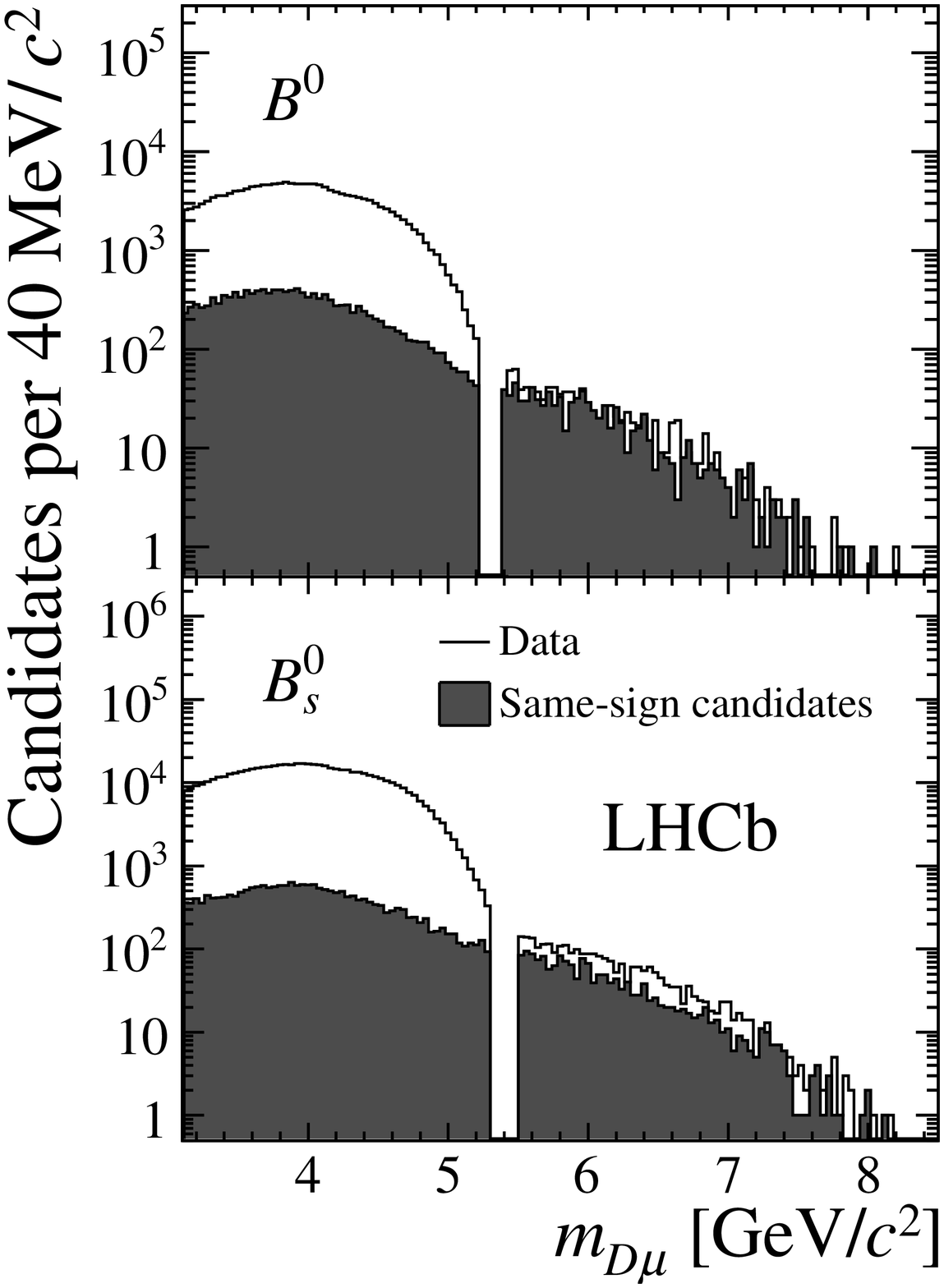}
\put(41,69){\includegraphics[width=0.17\textwidth]{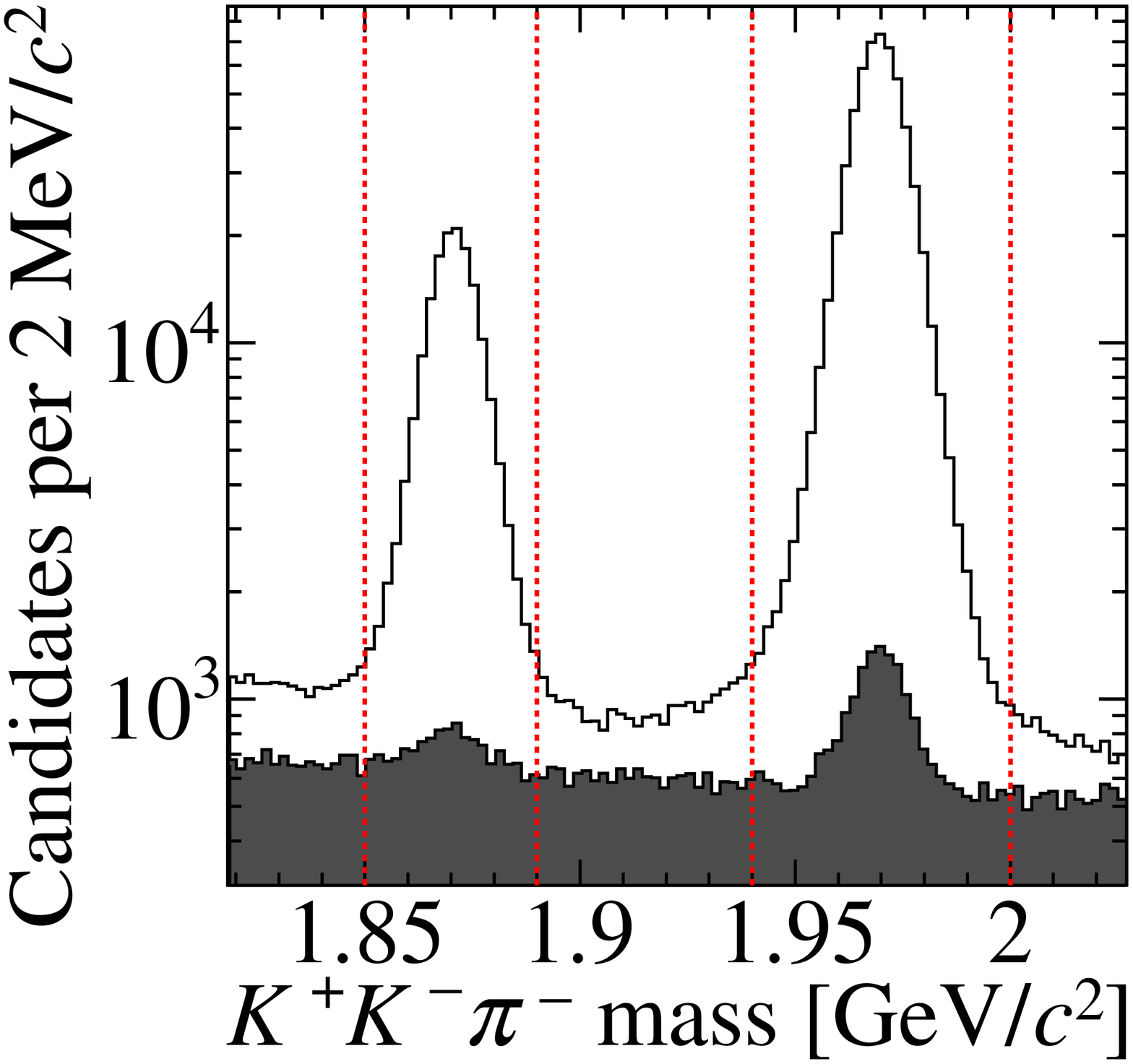}}
\end{overpic}\\
\caption{\label{fig:visibleMass} Distributions of $D\mu$ mass for (top panel) reference candidates, formed by combining $\Dm \to K^+K^-\pi^-$ candidates with $\mu^+$ candidates, and (bottom panel)  signal candidates formed by $\Dsm \to K^+K^-\pi^-$ candidates combined with $\mu^+$ candidates.  The inset shows the $K^+K^-\pi^-$-mass distribution with vertical lines enclosing the \Dm (\Dsm)  candidates used to form the reference (signal) candidates. The dark-filled histograms show same-sign candidate distributions.}
\end{figure}

A total of approximately 468\,000 (141\,000) signal (reference) candidates, formed by combining $K^+K^-\pi^-$ candidates in the \Dsm (\Dm) signal range with $\mu^+$ candidates, satisfy the selection. Figure~\ref{fig:visibleMass} shows the relevant mass distributions. The enhancements of the signal and reference distributions over the corresponding same-sign distributions for $m_{\D\mu}< 5.5 \gevcc$ are due to bottom-hadron decays. The absence of candidates at $m_{\D\mu}\approx 5.3 \gevcc$ results from the $B^0_{(s)} \to D^-_{(s)}\pi^+$ veto. The two peaks in the $K^+K^-\pi^-$ distributions of same-sign candidates are due to genuine charm decays accidentally combined with muon candidates. Along with $\Bs \to [K^+K^-\pi^-]_\DSSm \mu^+\nu_\mu$ decays, many \Bs decays potentially useful for the lifetime measurement contribute signal candidates, including decays into $D_{(s)}^{**}(\to \DSSm X) \mu^+ \nu_{\mu}$,  $\Dsm \tau^+ (\to \mu^+ \nu_{\mu} \bar{\nu}_\tau) \nu_\tau$, $\Dssm (\to \Dsm X) \tau^+ (\to \mu^+ \nu_\mu \bar{\nu}_{\tau}) \nu_{\tau}$, and $D^{**}_{s} (\to \DSSm X) \tau^+ (\to \mu^+\nu_\mu \bar{\nu}_{\tau}) \nu_\tau$ final states.\footnote{Here and in the following, the symbol $D^{**}_{(s)}$ identifies collectively higher orbital excitations of $D^{-}_{(s)}$ mesons.} Similarly, along with the $\Bz \to [K^+K^-\pi^-]_\DSm \mu^+\nu_\mu$ decays, potential reference candidates come from \Bz decays into $D^{**}(\to D^{\scalebox{0.4}{(}*\scalebox{0.4}{)}-} X ) \mu^+ \nu_\mu$,  $D^- \tau^+ (\to \mu^+ \nu_\mu \bar{\nu}_\tau) \nu_\tau$, $D^{*-} (\to \Dm X) \tau^+ (\to \mu^+ \nu_\mu \bar{\nu}_\tau) \nu_\tau$, and $D^{**}(\to D^{\scalebox{0.4}{(}*\scalebox{0.4}{)}-} X )\tau^+ (\to \mu^+ \nu_\mu \bar{\nu}_\tau) \nu_\tau$ final states. However, we restrict the signal (reference) decays solely to the $\Bs \to [K^+K^-\pi^-]_\DSSm \mu^+\nu_\mu$  ($\Bz \to [K^+K^-\pi^-]_\DSm \mu^+\nu_\mu$) channels because they contribute 95\% (91\%) of the inclusive $K^+K^-\pi^-\mu^+$ yield from semileptonic \Bz (\Bs) decays and require smaller and better-known $k$-factor corrections to relate the observed decay times to their true values.

\begin{figure}
\centering
\includegraphics[width=0.48\textwidth]{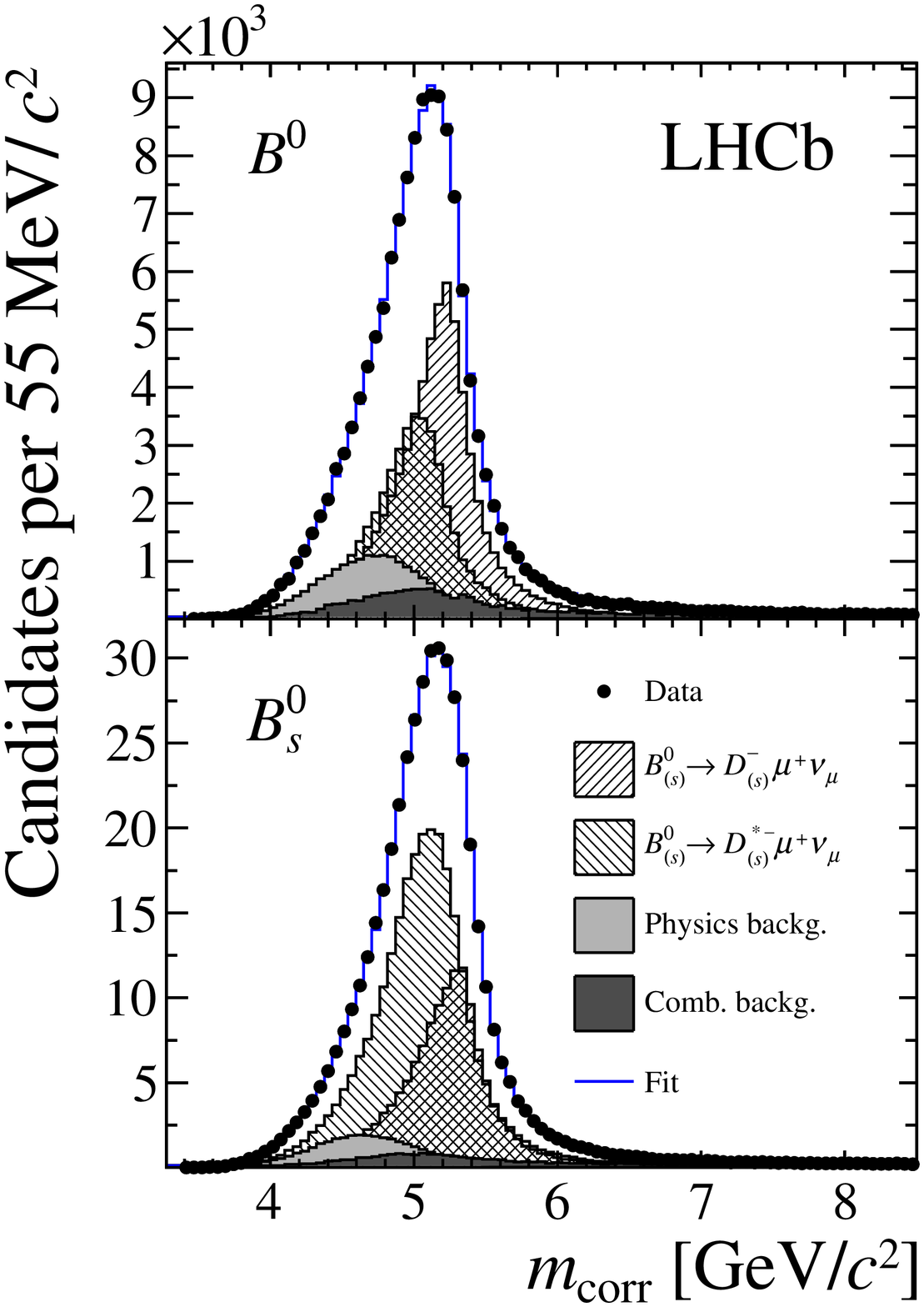}\\
\caption{Corrected-mass distributions for (top panel) reference $\Bz \to [K^+ K^- \pi^-]_{\DSm} \mu^+ \nu_\mu$ 
and (bottom panel) signal  $\Bs \to [K^+K^-\pi^-]_{\DSSm} \mu^+ \nu_\mu$ 
candidates satisfying the selection. Results of the global composition-fit are overlaid. In the \Bs fit projection, the lower- and higher-mass background components described in the text are displayed as a single, merged ``physics background" component. \label{fig:B_M_AfterSelection}} 
\end{figure}

A reliable understanding of the sample composition is essential for unbiased lifetime results.  An unbiased determination from simulation of the acceptances and mass distributions as functions of decay time requires that the simulated sample mirrors the data composition. We therefore weight the composition of the simulated samples according to the results of a least-squares fit to the $m_{\rm corr}$ distributions in data, shown in Fig.~\ref{fig:B_M_AfterSelection}. In the \Bs sample, such a global composition-fit includes the two signal components, $\Bs \to [K^+K^-\pi^-]_\Dsm \mu^+ \nu_\mu$ and $\Bs \to [K^+K^-\pi^-]_\Dssm \mu^+ \nu_\mu$;  a combinatorial component; and two physics backgrounds. The physics backgrounds are formed by grouping together contributions with similar corrected-mass distributions, determined from simulation. They are divided into contributions at lower values of corrected mass ($\Bz \to D^{\scalebox{0.4}{(}*\scalebox{0.4}{)}-}D^{\scalebox{0.4}{(}*\scalebox{0.4}{)}+}_s$, $\Bp \to \Dbar{}^{\scalebox{0.4}{(}*\scalebox{0.4}{)}0}D^{\scalebox{0.4}{(}*\scalebox{0.4}{)}+}_s$, and $D^{**}(\to \DSSm X) \mu^+ \nu_\mu$) and at higher corrected-mass values ($\Bp \to \DSSm K^+ \mu^+  \nu_\mu X$, $\Bz \to \DSSm \Kz \mu^+ \nu_\mu X$, and $\Bs \to \Dsm \tau^+ (\to \mu^+ \nu_\mu \bar{\nu}_\tau) \nu_\tau X$). The distributions of all components are modeled empirically from simulation, except for the combinatorial component, which is modeled using same-sign data. Contributions expected to be smaller than 0.5\% are neglected. The effect of this approximation and of possible variations of the relative proportions within each fit category are treated as contributions to the systematic uncertainties. The fit $p$-value is 62.1\% and the fractions of each component are determined with absolute statistical uncertainties in the range 0.13\%--0.91\%. A simpler composition fit is used for the \Bz sample. Signal and combinatorial components are chosen similarly to the \Bs case; the contributions from $\Bz \to D^{**-}(\to D^{\scalebox{0.4}{(}*\scalebox{0.4}{)}-} X )\mu^+ \nu_\mu$ and $\Bp \to \Dm \mu^+ \nu_\mu X$ decays have sufficiently similar distributions to be merged into a single physics-background component. The results of the corrected-mass fit of the reference sample also offer a validation of the approach, since the composition of this sample is known precisely from other experiments. The largest discrepancy observed among the individual fractional contributions is 1.3 statistical standard deviations.

\begin{figure}[t]
\centering
\includegraphics[width=0.48\textwidth]{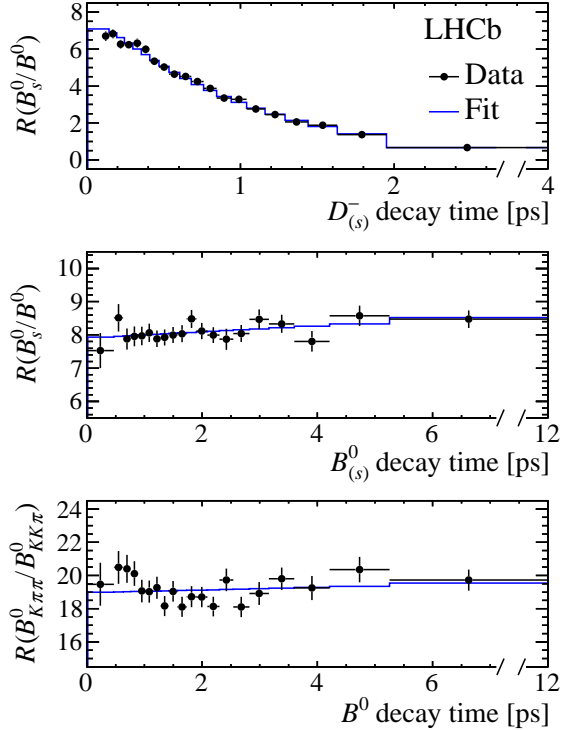}\\
\caption{Ratio between acceptance-corrected yields of signal  $\Bs \to [K^+K^-\pi^-]_{\DSSm} \mu^+\nu_\mu$ 
and reference $\Bz \to [K^+K^-\pi^-]_{\DSm} \mu^+\nu_\mu$  decay yields as a function of (top panel) charm-meson and (middle panel) bottom-meson decay time. The bottom panel shows the ratio between acceptance-corrected \Bz decay yields in the $[K^+\pi^-\pi^-]_{\DSm} \mu^+\nu_\mu$  and $[K^+K^-\pi^-]_{\DSm} \mu^+\nu_\mu$ channels as a function of \Bz decay time.  Fit results are overlaid. Relevant for the results is only the slope of the ratios as a function of decay time; absolute ratios, which depend on the decay yields, weighting, and efficiencies, are irrelevant. \label{fig:timefit_Bdratio}}
\end{figure}

The composition fit is sufficient for the determination of $\Delta_\Gamma(D)$, where no $k$-factor corrections are needed since the final state is fully reconstructed. We determine $\Delta_\Gamma(D)$ through a least-squares fit of the ratio of signal \Bs and reference \Bz yields as a function of the charm-meson decay time in the range 0.1--4.0\,ps. The yields of signal $\Bs \to [K^+K^-\pi^-]_\DSSm \mu^+\nu_\mu$ and reference $\Bz \to [K^+K^-\pi^-]_\DSm \mu^+\nu_\mu$ decays are determined in each of 20 decay-time bins with a $m_{\rm corr}$ fit similar to the global composition-fit. The two signal and the two physics-background contributions are each merged into a single component according to the total proportions determined by the global fit and their decay-time dependence as determined from simulation. The fit includes the decay-time resolution and the ratio between signal and reference decay-time acceptances, which is determined from simulation to be uniform within 1\%.  The fit is shown in the top panel of Fig.~\ref{fig:timefit_Bdratio}; it has 34\% $p$-value and determines $\Delta_\Gamma(D) = 1.0131 \pm 0.0117$\invps.

The measurement of $\Delta_\Gamma(B)$ requires an acceptance correction for the differences between signal and reference decays and the $k$-factor correction. The acceptance correction accounts for the difference in decay-time-dependent efficiency due to the combined effect of the difference between \Dm and \Dsm lifetimes and the online requirements on the spatial separation between $D^-_{(s)}$ and $B^0_{(s)}$ decay vertices: we apply to the \Bs sample a per-candidate weight, $w_i \equiv \exp[\Delta_\Gamma(D) t(\Dsm)]$, based on the $\Delta_\Gamma(D)$ result and the \Dsm decay time, such that the \Dsm and \Dm decay-time distributions become consistent. The $k$-factor correction is a candidate-specific correction, where the average missing momentum in a simulated sample is used to correct the reconstructed momentum in data.  The $k$-factor dependence on the kinematic properties of each candidate is included through a dependence on $m_{D\mu}$,  $k(m_{D\mu}) = \left\langle p_{D\mu}/p_{\rm true}\right\rangle$, where $p_{\rm true}$ indicates the true momentum of the $B^0_{(s)}$ meson. The equalization of the compositions of simulated and experimental data samples ensures that the $k$-factor distribution specific to each of the four signal and reference decays is unbiased. We determine $\Delta_\Gamma(B)$ with the same fit of $m_{\rm corr}$ used to measure $\Delta_\Gamma(D)$ but where the ratios of signal and reference yields are determined as functions of the $B^0_{(s)}$ decay time. The decay-time smearing due to the $k$-factor spread is included in the fit.  After the \Dsm lifetime weighting, the decay-time acceptances of simulated signal and reference modes are consistent, with a $p$-value of $83\%$, and are not included in the fit. The fit is shown in the middle panel of Fig.~\ref{fig:timefit_Bdratio}; the resulting width difference is $\Delta_\Gamma(B) =  -0.0115 \pm 0.0053$\invps, with 91\% $p$-value.

To check against biases due to differing acceptances and kinematic properties, the analysis is validated with a null test. We repeat the width-difference determination by using the same reference $\Bz \to [K^+K^-\pi^-]_\DSm \mu^+\nu_\mu$ sample and replacing the signal decays with 2.1 million $\Bz \to [K^+\pi^-\pi^-]_\DSm \mu^+\nu_\mu$ decays, where the \Dm is reconstructed in the $K^+\pi^-\pi^-$ final state (Fig.~\ref{fig:timefit_Bdratio}, bottom panel). Differing momentum and vertex-displacement selection criteria induce up to 10\% differences between acceptances as a function of $\Dm$ decay time and up to 25\% variations as a function of \Bz decay time. Acceptance ratios are therefore included in the fit. The $p$-values are 21\% for the \Bz fit and 33\% for the \Dm fit. The resulting width differences, $\Delta_\Gamma(D) = (-19 \pm 10)\times 10^{-3}$\invps and $\Delta_\Gamma(B) = (-4.1 \pm 5.4)\times 10^{-3}$\invps, are consistent with zero.

We assess independent systematic uncertainties due to (i) potential fit biases; (ii) assumptions on the components contributing to the sample and their mass distributions; (iii) assumptions on the signal decay model, e.g., choice of $\Bs \to \Dssm$ form factors; (iv) uncertainties on the decay-time acceptances; (v) uncertainties on the decay-time resolution; (vi) contamination from \Bs candidates produced in $B_c^+$ decays; and (vii) mismodeling of transverse-momentum (\pt) differences between \Bz and \Bs mesons.  We evaluate each contribution by including the relevant effect in the model and repeating the whole analysis on ensembles of simulated experiments that mirror the data. For the $\Delta_\Gamma(D)$ result, the systematic uncertainty is dominated by a 0.0049\invps contribution due to the decay-time acceptance, and a 0.0039\invps contribution due to the decay-time resolution. A smaller contribution of 0.0018\invps arises from possible mismodeling of \pt differences in \Bz and \Bs production. For the $\Delta_\Gamma(B)$ result,  a 0.0028\invps uncertainty from mismodeling of \pt differences between \Bz and \Bs mesons and a 0.0025\invps contribution from the \Bs decay model dominate. Smaller contributions arise from \Bc feed-down (0.0010\invps), residual fit biases (0.0009\invps), sample composition (0.0005\invps), and decay-time acceptance and resolution (0.0004\invps each). The uncertainties associated with the limited size of simulated samples are included in the fit $\chi^2$  and contribute up to 20\% of the statistical uncertainties. The uncertainty in the decay length has negligible impact. Consistency checks based on repeating the measurement independently on subsamples chosen according to data-taking time, online-selection criteria, charged-particle and vertex multiplicities, momentum of the $K^+K^-\pi^-\mu^+$ system, and whether only the $\Dsm \mu^+ \nu_\mu$ or the $\Dssm \mu^+ \nu_\mu$ channel is considered as signal, all yield results compatible with statistical fluctuations.

In summary, we report world-leading measurements of \Bs and \Dsm meson lifetimes using a novel method. We reconstruct $\Bs \to \Dssm \mu^+\nu_\mu$ and $\Bs \to \Dsm \mu^+\nu_\mu$ decays from proton-proton collision data collected by the LHCb experiment and corresponding to 3.0\invfb of integrated luminosity. We use $\Bz \to \Dstarm \mu^+\nu_\mu$ and $\Bz \to \Dm \mu^+\nu_\mu$ decays reconstructed in the same final state as a reference to suppress systematic uncertainties. The resulting width differences are $\Delta_\Gamma(B) = \dB\pm \eStatdB\stat \pm \eSystdB\syst$\invps and $\Delta_\Gamma(D) = \dD\pm \eStatdD\stat \pm \eSystdD\syst$\invps. Their correlation is negligible. Using the known values of the $B^0$~\cite{PDG2016, Aaij:2014owa} and $D^-$ lifetimes~\cite{PDG2016,Link:2002bx}, we determine the flavor-specific \Bs lifetime, $\taufs = \tB \pm \eStattB\stat \pm \eSysttB\syst\pm\eReftB\refB$\,ps, and the \Dsm lifetime, $\tau_{\Dsm} = \tD \pm \eStattD\stat \pm \eSysttD\syst\pm \eReftD\refD$\,ps; the last uncertainties are due to the limited knowledge of the \Bz and \Dm lifetime, respectively. The results are consistent with, and significantly more precise than the current values~\cite{Link:2005ew,Abazov:2014rua,LHCb-PAPER-2014-037}. They might offer improved insight into the interplay between strong and weak interactions in the dynamics of heavy mesons and sharpen the reach of current and future indirect searches for non-standard-model physics.

\section*{Acknowledgments}

\noindent We thank Alexander Lenz for useful discussions. We express our gratitude to our colleagues in the CERN
accelerator departments for the excellent performance of the LHC. We
thank the technical and administrative staff at the LHCb
institutes. We acknowledge support from CERN and from the national
agencies: CAPES, CNPq, FAPERJ and FINEP (Brazil); MOST and NSFC (China);
CNRS/IN2P3 (France); BMBF, DFG and MPG (Germany); INFN (Italy); 
NWO (The Netherlands); MNiSW and NCN (Poland); MEN/IFA (Romania); 
MinES and FASO (Russia); MinECo (Spain); SNSF and SER (Switzerland); 
NASU (Ukraine); STFC (United Kingdom); NSF (USA).
We acknowledge the computing resources that are provided by CERN, IN2P3 (France), KIT and DESY (Germany), INFN (Italy), SURF (The Netherlands), PIC (Spain), GridPP (United Kingdom), RRCKI and Yandex LLC (Russia), CSCS (Switzerland), IFIN-HH (Romania), CBPF (Brazil), PL-GRID (Poland) and OSC (USA). We are indebted to the communities behind the multiple open 
source software packages on which we depend.
Individual groups or members have received support from AvH Foundation (Germany),
EPLANET, Marie Sk\l{}odowska-Curie Actions and ERC (European Union), 
Conseil G\'{e}n\'{e}ral de Haute-Savoie, Labex ENIGMASS and OCEVU, 
R\'{e}gion Auvergne (France), RFBR and Yandex LLC (Russia), GVA, XuntaGal and GENCAT (Spain), Herchel Smith Fund, The Royal Society, Royal Commission for the Exhibition of 1851 and the Leverhulme Trust (United Kingdom).

\addcontentsline{toc}{section}{References}
\setboolean{inbibliography}{true}
%\bibliographystyle{LHCb}
%\bibliography{main,LHCb-PAPER,LHCb-CONF,LHCb-DP,LHCb-TDR}

\begin{mcitethebibliography}{10}
\mciteSetBstSublistMode{n}
\mciteSetBstMaxWidthForm{subitem}{\alph{mcitesubitemcount})}
\mciteSetBstSublistLabelBeginEnd{\mcitemaxwidthsubitemform\space}
{\relax}{\relax}

\bibitem{Lenz:2014jha}
For a recent review, see A.~Lenz, \ifthenelse{\boolean{articletitles}}{\emph{{Lifetimes and HQE}},
  }{}\href{http://arxiv.org/abs/1405.3601}{{\normalfont\ttfamily
  arXiv:1405.3601}} and references therein\relax
\mciteBstWouldAddEndPuncttrue
\mciteSetBstMidEndSepPunct{\mcitedefaultmidpunct}
{\mcitedefaultendpunct}{\mcitedefaultseppunct}\relax
\EndOfBibitem
%\bibitem{Shifman:1986sm}
%M.~A. Shifman and M.~B. Voloshin,
%  \ifthenelse{\boolean{articletitles}}{\emph{{On annihilation of mesons built
%  from heavy and light quark and \Bz-\Bzb oscillations}}, }{}Sov.\ J.\ Nucl.\
%  Phys.\  \textbf{45} (1987) 292\relax
%\mciteBstWouldAddEndPuncttrue
%\mciteSetBstMidEndSepPunct{\mcitedefaultmidpunct}
%{\mcitedefaultendpunct}{\mcitedefaultseppunct}\relax
%\EndOfBibitem
%\bibitem{Shifman:1987rj}
%M.~A. Shifman and M.~B. Voloshin,
%  \ifthenelse{\boolean{articletitles}}{\emph{{On production of D and $D^*$
%  mesons in B meson decays}}, }{}Sov.\ J.\ Nucl.\ Phys.\  \textbf{47} (1988)
%  511\relax
%\mciteBstWouldAddEndPuncttrue
%\mciteSetBstMidEndSepPunct{\mcitedefaultmidpunct}
%{\mcitedefaultendpunct}{\mcitedefaultseppunct}\relax
%\EndOfBibitem
%\bibitem{Eichten:1989zv}
%E.~Eichten and B.~R. Hill, \ifthenelse{\boolean{articletitles}}{\emph{{An
%  effective field theory for the calculation of matrix elements involving heavy
%  quarks}}, }{}\href{http://dx.doi.org/10.1016/0370-2693(90)92049-O}{Phys.\
%  Lett.\  \textbf{B234} (1990) 511}\relax
%\mciteBstWouldAddEndPuncttrue
%\mciteSetBstMidEndSepPunct{\mcitedefaultmidpunct}
%{\mcitedefaultendpunct}{\mcitedefaultseppunct}\relax
%\EndOfBibitem
%\bibitem{Guberina:1979fe}
%B.~Guberina, R.~D. Peccei, and R.~R{\"u}ckl,
%  \ifthenelse{\boolean{articletitles}}{\emph{{Weak decays of heavy quarks}},
%  }{}\href{http://dx.doi.org/10.1016/0370-2693(80)90674-7}{Phys.\ Lett.\
%  \textbf{B91} (1980) 116}\relax
%\mciteBstWouldAddEndPuncttrue
%\mciteSetBstMidEndSepPunct{\mcitedefaultmidpunct}
%{\mcitedefaultendpunct}{\mcitedefaultseppunct}\relax
%\EndOfBibitem
%\bibitem{Guberina:1979re}
%B.~Guberina, R.~D. Peccei, and R.~R{\"u}ckl,
%  \ifthenelse{\boolean{articletitles}}{\emph{{Effects of QCD on the Cabibbo
%  patterns in $B$ meson decays}},
%  }{}\href{http://dx.doi.org/10.1016/0370-2693(80)90077-5}{Phys.\ Lett.\
%  \textbf{B90} (1980) 169}\relax
%\mciteBstWouldAddEndPuncttrue
%\mciteSetBstMidEndSepPunct{\mcitedefaultmidpunct}
%{\mcitedefaultendpunct}{\mcitedefaultseppunct}\relax
%\EndOfBibitem
%\bibitem{Isgur:1989vq}
%N.~Isgur and M.~B. Wise, \ifthenelse{\boolean{articletitles}}{\emph{{Weak
%  decays of heavy mesons in the static quark approximation}},
%  }{}\href{http://dx.doi.org/10.1016/0370-2693(89)90566-2}{Phys.\ Lett.\
%  \textbf{B232} (1989) 113}\relax
%\mciteBstWouldAddEndPuncttrue
%\mciteSetBstMidEndSepPunct{\mcitedefaultmidpunct}
%{\mcitedefaultendpunct}{\mcitedefaultseppunct}\relax
%\EndOfBibitem
%\bibitem{Isgur:1989ed}
%N.~Isgur and M.~B. Wise, \ifthenelse{\boolean{articletitles}}{\emph{{Weak
%  transition form-factors between heavy mesons}},
%  }{}\href{http://dx.doi.org/10.1016/0370-2693(90)91219-2}{Phys.\ Lett.\
%  \textbf{B237} (1990) 527}\relax
%\mciteBstWouldAddEndPuncttrue
%\mciteSetBstMidEndSepPunct{\mcitedefaultmidpunct}
%{\mcitedefaultendpunct}{\mcitedefaultseppunct}\relax
%\EndOfBibitem
%\bibitem{Grinstein:1990mj}
%B.~Grinstein, \ifthenelse{\boolean{articletitles}}{\emph{{The static quark
%  effective theory}},
%  }{}\href{http://dx.doi.org/10.1016/0550-3213(90)90349-I}{Nucl.\ Phys.\
%  \textbf{B339} (1990) 253}\relax
%\mciteBstWouldAddEndPuncttrue
%\mciteSetBstMidEndSepPunct{\mcitedefaultmidpunct}
%{\mcitedefaultendpunct}{\mcitedefaultseppunct}\relax
%\EndOfBibitem
%\bibitem{Falk:1990yz}
%A.~F. Falk, H.~Georgi, B.~Grinstein, and M.~B. Wise,
%  \ifthenelse{\boolean{articletitles}}{\emph{{Heavy meson form-factors from
%  QCD}}, }{}\href{http://dx.doi.org/10.1016/0550-3213(90)90591-Z}{Nucl.\ Phys.\
%   \textbf{B343} (1990) 1}\relax
%\mciteBstWouldAddEndPuncttrue
%\mciteSetBstMidEndSepPunct{\mcitedefaultmidpunct}
%{\mcitedefaultendpunct}{\mcitedefaultseppunct}\relax
%\EndOfBibitem
%\bibitem{Georgi:1990um}
%H.~Georgi, \ifthenelse{\boolean{articletitles}}{\emph{{An effective field
%  theory for heavy quarks at low-energies}},
%  }{}\href{http://dx.doi.org/10.1016/0370-2693(90)91128-X}{Phys.\ Lett.\
%  \textbf{B240} (1990) 447}\relax
%\mciteBstWouldAddEndPuncttrue
%\mciteSetBstMidEndSepPunct{\mcitedefaultmidpunct}
%{\mcitedefaultendpunct}{\mcitedefaultseppunct}\relax
%\EndOfBibitem
\bibitem{HFAG}
Heavy Flavor Averaging Group, Y.~Amhis {\em et~al.},
  \ifthenelse{\boolean{articletitles}}{\emph{{Averages of $b$-hadron,
  $c$-hadron, and $\tau$-lepton properties as of summer 2016}},
  }{}\href{http://arxiv.org/abs/1612.07233}{{\normalfont\ttfamily
  arXiv:1612.07233}}, {updated results and plots available at
  \href{http://www.slac.stanford.edu/xorg/hfag/}{{\texttt{http://www.slac.stanford.edu/xorg/hfag/}}}}\relax
\mciteBstWouldAddEndPuncttrue
\mciteSetBstMidEndSepPunct{\mcitedefaultmidpunct}
{\mcitedefaultendpunct}{\mcitedefaultseppunct}\relax
\EndOfBibitem
\bibitem{Hartkorn:1999ga}
K.~Hartkorn and H.~G. Moser, \ifthenelse{\boolean{articletitles}}{\emph{{A new
  method of measuring $\Delta(\Gamma)/\Gamma$ in the \Bs-\Bsb system}},
  }{}\href{http://dx.doi.org/10.1007/s100520050472}{Eur.\ Phys.\ J.\
  \textbf{C8} (1999) 381}\relax
\mciteBstWouldAddEndPuncttrue
\mciteSetBstMidEndSepPunct{\mcitedefaultmidpunct}
{\mcitedefaultendpunct}{\mcitedefaultseppunct}\relax
\EndOfBibitem
\bibitem{LHCb-PAPER-2014-037}
LHCb collaboration, R.~Aaij {\em et~al.},
  \ifthenelse{\boolean{articletitles}}{\emph{{Measurement of the \Bsb meson
  lifetime in \Dsp \pim decays}},
  }{}\href{http://dx.doi.org/10.1103/PhysRevLett.113.172001}{Phys.\ Rev.\
  Lett.\  \textbf{113} (2014) 172001},
  \href{http://arxiv.org/abs/1407.5873}{{\normalfont\ttfamily
  arXiv:1407.5873}}\relax
\mciteBstWouldAddEndPuncttrue
\mciteSetBstMidEndSepPunct{\mcitedefaultmidpunct}
{\mcitedefaultendpunct}{\mcitedefaultseppunct}\relax
\EndOfBibitem
\bibitem{Abazov:2014rua}
D0 collaboration, V.~M. Abazov {\em et~al.},
  \ifthenelse{\boolean{articletitles}}{\emph{{Measurement of the $B_s^0$
  lifetime in the flavor-specific decay channel $B_s^0 \to D_s^- \mu^+\nu X$}},
  }{}\href{http://dx.doi.org/10.1103/PhysRevLett.114.062001}{Phys.\ Rev.\
  Lett.\  \textbf{114} (2015) 062001},
  \href{http://arxiv.org/abs/1410.1568}{{\normalfont\ttfamily
  arXiv:1410.1568}}\relax
\mciteBstWouldAddEndPuncttrue
\mciteSetBstMidEndSepPunct{\mcitedefaultmidpunct}
{\mcitedefaultendpunct}{\mcitedefaultseppunct}\relax
\EndOfBibitem
\bibitem{Link:2005ew}
FOCUS collaboration, J.~M. Link {\em et~al.},
  \ifthenelse{\boolean{articletitles}}{\emph{{A measurement of the \Dsp
  lifetime}}, }{}\href{http://dx.doi.org/10.1103/PhysRevLett.95.052003}{Phys.\
  Rev.\ Lett.\  \textbf{95} (2005) 052003},
  \href{http://arxiv.org/abs/hep-ex/0504056}{{\normalfont\ttfamily
  arXiv:hep-ex/0504056}}\relax
\mciteBstWouldAddEndPuncttrue
\mciteSetBstMidEndSepPunct{\mcitedefaultmidpunct}
{\mcitedefaultendpunct}{\mcitedefaultseppunct}\relax
\EndOfBibitem
\bibitem{Kodama:1991ij}
Fermilab E653 collaboration, K.~Kodama {\em et~al.},
  \ifthenelse{\boolean{articletitles}}{\emph{{Measurement of the relative
  branching fraction $\Gamma (D^0 \rightarrow K \mu \nu) / \Gamma (D^0
  \rightarrow \mu X)$}},
  }{}\href{http://dx.doi.org/10.1103/PhysRevLett.66.1819}{Phys.\ Rev.\ Lett.\
  \textbf{66} (1991) 1819}\relax
\mciteBstWouldAddEndPuncttrue
\mciteSetBstMidEndSepPunct{\mcitedefaultmidpunct}
{\mcitedefaultendpunct}{\mcitedefaultseppunct}\relax
\EndOfBibitem
\bibitem{PDG2016}
Particle Data Group, C.~Patrignani {\em et~al.},
  \ifthenelse{\boolean{articletitles}}{\emph{{\href{http://pdg.lbl.gov/}{Review
  of particle physics}}},
  }{}\href{http://dx.doi.org/10.1088/1674-1137/40/10/100001}{Chin.\ Phys.\
  \textbf{C40} (2016) 100001}\relax
\mciteBstWouldAddEndPuncttrue
\mciteSetBstMidEndSepPunct{\mcitedefaultmidpunct}
{\mcitedefaultendpunct}{\mcitedefaultseppunct}\relax
\EndOfBibitem
\bibitem{Abulencia:2006ze}
CDF collaboration, A.~Abulencia {\em et~al.},
  \ifthenelse{\boolean{articletitles}}{\emph{{Observation of \Bs-\Bsb
  oscillations}},
  }{}\href{http://dx.doi.org/10.1103/PhysRevLett.97.242003}{Phys.\ Rev.\ Lett.\
   \textbf{97} (2006) 242003},
  \href{http://arxiv.org/abs/hep-ex/0609040}{{\normalfont\ttfamily
  arXiv:hep-ex/0609040}}\relax
\mciteBstWouldAddEndPuncttrue
\mciteSetBstMidEndSepPunct{\mcitedefaultmidpunct}
{\mcitedefaultendpunct}{\mcitedefaultseppunct}\relax
\EndOfBibitem
\bibitem{Leonardo:2006fq}
N.~T. Leonardo, {\em {Analysis of \Bs flavor oscillations at CDF}}, PhD thesis,
  FERMILAB-THESIS-2006-18, (2006)\relax
\mciteBstWouldAddEndPuncttrue
\mciteSetBstMidEndSepPunct{\mcitedefaultmidpunct}
{\mcitedefaultendpunct}{\mcitedefaultseppunct}\relax
\EndOfBibitem
\bibitem{Alves:2008zz}
LHCb collaboration, A.~A. Alves~Jr.\ {\em et~al.},
  \ifthenelse{\boolean{articletitles}}{\emph{{The \lhcb detector at the LHC}},
  }{}\href{http://dx.doi.org/10.1088/1748-0221/3/08/S08005}{JINST \textbf{3}
  (2008) S08005}\relax
\mciteBstWouldAddEndPuncttrue
\mciteSetBstMidEndSepPunct{\mcitedefaultmidpunct}
{\mcitedefaultendpunct}{\mcitedefaultseppunct}\relax
\EndOfBibitem
\bibitem{Aaij:2014jba}
LHCb collaboration, R.~Aaij {\em et~al.},
  \ifthenelse{\boolean{articletitles}}{\emph{{LHCb detector performance}},
  }{}\href{http://dx.doi.org/10.1142/S0217751X15300227}{Int.\ J.\ Mod.\ Phys.\
  \textbf{A30} (2015) 1530022},
  \href{http://arxiv.org/abs/1412.6352}{{\normalfont\ttfamily
  arXiv:1412.6352}}\relax
\mciteBstWouldAddEndPuncttrue
\mciteSetBstMidEndSepPunct{\mcitedefaultmidpunct}
{\mcitedefaultendpunct}{\mcitedefaultseppunct}\relax
\EndOfBibitem
\bibitem{LHCb-PROC-2011-006}
M.~Clemencic {\em et~al.}, \ifthenelse{\boolean{articletitles}}{\emph{{The
  \lhcb simulation application, Gauss: Design, evolution and experience}},
  }{}\href{http://dx.doi.org/10.1088/1742-6596/331/3/032023}{{J.\ Phys.\ Conf.\
  Ser.\ } \textbf{331} (2011) 032023}\relax
\mciteBstWouldAddEndPuncttrue
\mciteSetBstMidEndSepPunct{\mcitedefaultmidpunct}
{\mcitedefaultendpunct}{\mcitedefaultseppunct}\relax
\EndOfBibitem
\bibitem{LHCb-PROC-2010-056}
I.~Belyaev {\em et~al.}, \ifthenelse{\boolean{articletitles}}{\emph{{Handling
  of the generation of primary events in Gauss, the LHCb simulation
  framework}}, }{}\href{http://dx.doi.org/10.1088/1742-6596/331/3/032047}{{J.\
  Phys.\ Conf.\ Ser.\ } \textbf{331} (2011) 032047}\relax
\mciteBstWouldAddEndPuncttrue
\mciteSetBstMidEndSepPunct{\mcitedefaultmidpunct}
{\mcitedefaultendpunct}{\mcitedefaultseppunct}\relax
\EndOfBibitem
\bibitem{Caprini:1997mu}
I.~Caprini, L.~Lellouch, and M.~Neubert,
  \ifthenelse{\boolean{articletitles}}{\emph{{Dispersive bounds on the shape of
  $\Bz \to D^{(*)} \ell \bar{\nu}_{\ell}$ form factors}},
  }{}\href{http://dx.doi.org/10.1016/S0550-3213(98)00350-2}{Nucl.\ Phys.\
  \textbf{B530} (1998) 153},
  \href{http://arxiv.org/abs/hep-ph/9712417}{{\normalfont\ttfamily
  arXiv:hep-ph/9712417}}\relax
\mciteBstWouldAddEndPuncttrue
\mciteSetBstMidEndSepPunct{\mcitedefaultmidpunct}
{\mcitedefaultendpunct}{\mcitedefaultseppunct}\relax
\EndOfBibitem
\bibitem{Aaij:2012me}
R.~Aaij {\em et~al.}, \ifthenelse{\boolean{articletitles}}{\emph{{The LHCb
  trigger and its performance in 2011}},
  }{}\href{http://dx.doi.org/10.1088/1748-0221/8/04/P04022}{JINST \textbf{8}
  (2013) P04022}, \href{http://arxiv.org/abs/1211.3055}{{\normalfont\ttfamily
  arXiv:1211.3055}}\relax
\mciteBstWouldAddEndPuncttrue
\mciteSetBstMidEndSepPunct{\mcitedefaultmidpunct}
{\mcitedefaultendpunct}{\mcitedefaultseppunct}\relax
\EndOfBibitem
\bibitem{Aaij:2014owa}
LHCb collaboration, R.~Aaij {\em et~al.},
  \ifthenelse{\boolean{articletitles}}{\emph{{Measurements of the $B^+, B^0,
  B^0_s$ meson and $\Lambda^0_b$ baryon lifetimes}},
  }{}\href{http://dx.doi.org/10.1007/JHEP04(2014)114}{JHEP \textbf{04} (2014)
  114}, \href{http://arxiv.org/abs/1402.2554}{{\normalfont\ttfamily
  arXiv:1402.2554}}\relax
\mciteBstWouldAddEndPuncttrue
\mciteSetBstMidEndSepPunct{\mcitedefaultmidpunct}
{\mcitedefaultendpunct}{\mcitedefaultseppunct}\relax
\EndOfBibitem
\bibitem{Link:2002bx}
FOCUS collaboration, J.~M. Link {\em et~al.},
  \ifthenelse{\boolean{articletitles}}{\emph{{New measurements of the $D^0$ and
  $D^+$ lifetimes}},
  }{}\href{http://dx.doi.org/10.1016/S0370-2693(02)01934-2}{Phys.\ Lett.\
  \textbf{B537} (2002) 192},
  \href{http://arxiv.org/abs/hep-ex/0203037}{{\normalfont\ttfamily
  arXiv:hep-ex/0203037}}\relax
\mciteBstWouldAddEndPuncttrue
\mciteSetBstMidEndSepPunct{\mcitedefaultmidpunct}
{\mcitedefaultendpunct}{\mcitedefaultseppunct}\relax
\EndOfBibitem
\end{mcitethebibliography}
\ifx\mcitethebibliography\mciteundefinedmacro
\PackageError{LHCb.bst}{mciteplus.sty has not been loaded}
{This bibstyle requires the use of the mciteplus package.}\fi
\providecommand{\href}[2]{#2}

% This should be taken out in the final paper
%\newpage
%\input{PRLjustification}
%\input{supplementary}
%\input{wc}

% Author List ----------------------------                                                                                                                                                                                                                                                                                                
\newpage
\centerline{\large\bf LHCb collaboration}
\begin{flushleft}
\small
R.~Aaij$^{40}$,
B.~Adeva$^{39}$,
M.~Adinolfi$^{48}$,
Z.~Ajaltouni$^{5}$,
S.~Akar$^{59}$,
J.~Albrecht$^{10}$,
F.~Alessio$^{40}$,
M.~Alexander$^{53}$,
S.~Ali$^{43}$,
G.~Alkhazov$^{31}$
% ...

{\footnotesize \it
% ...
$ ^{5}$Clermont Universit{\'e}, Universit{\'e} Blaise Pascal, CNRS/IN2P3, LPC, Clermont-Ferrand, France\\
% ...
$ ^{10}$Fakult{\"a}t Physik, Technische Universit{\"a}t Dortmund, Dortmund, Germany\\
% ...
$ ^{31}$Petersburg Nuclear Physics Institute (PNPI), Gatchina, Russia\\
% ...
$ ^{39}$Universidad de Santiago de Compostela, Santiago de Compostela, Spain\\
$ ^{40}$European Organization for Nuclear Research (CERN), Geneva, Switzerland\\
% ...
$ ^{43}$Nikhef National Institute for Subatomic Physics, Amsterdam, The Netherlands\\
% ...
$ ^{48}$H.H. Wills Physics Laboratory, University of Bristol, Bristol, United Kingdom\\
% ...
$ ^{53}$School of Physics and Astronomy, University of Glasgow, Glasgow, United Kingdom\\
% ...
$ ^{59}$University of Cincinnati, Cincinnati, OH, United States\\
% ...
}
\end{flushleft}

\end{document}